\documentstyle[psfig]{aipproc}

\def\gray{{$\gamma$-ray}}
\def\grays{{$\gamma$-rays}}

\def\fun#1#2{\lower3.6pt\vbox{\baselineskip0pt\lineskip.9pt
  \ialign{$\mathsurround=0pt#1\hfil##\hfil$\crcr#2\crcr\sim\crcr}}}

\begin{document}


\title{The Extragalactic Gamma-Ray Background}
   
\author{F.W. Stecker}

\address{Laboratory for High Energy Astrophysics, Code 661,
NASA/Goddard Space Flight Center, Greenbelt, MD 20771, USA.}

\author{M.H. Salamon}
\address{Physics Department, University of Utah, Salt Lake City, UT 84112, USA}

\vspace{0.8cm}

\maketitle{}

\begin{abstract}

The {\it COMPTEL} and {\it {\it EGRET}} detectors aboard the Compton Gamma-Ray
Observatory measured an extragalactic $\gamma$-ray background (EGRB)
extending from
$\sim$ 1 MeV to $\sim$ 100 GeV. Calculations performed making reasonable 
assumptions indicate that blazars can account for the background between 
$\sim$ 10 MeV and $\sim$ 10 GeV. Below 30 MeV, the background flux and 
spectrum are not very well determined and a dedicated satellite detector will 
be required to remedy this situation. Below 10 MeV, supernovae and possibly 
AGN may contribute to the extragalactic background flux. Above 10 GeV, the 
role of blazars in contributing to the background is unclear because we do not 
have data on their spectra at these energies and because theoretical models 
predict that many of them will have spectra which should cut off in this 
energy range. At these higher energies, a new component, perhaps from 
topological defects, may contribute to the background, as well as X-ray 
selected BL Lac objects. {\it {\it GLAST}} should provide important data on 
the emission of extragalactic sources above 10 GeV and help resolve this issue.
{\it {\it GLAST}} may also be able to detect the signature of intergalactic
absorption by pair production interactions of background $\gamma$-rays of 
energy above $\sim$ 20 GeV with starlight photons, this signature being a
steepening of the background spectrum.

\end{abstract}

\section{Introduction}

The EGRB measured by {\it {\it EGRET}} can be represented as of the power-law
form

\begin{equation}
\frac{dN_{\gamma}}{dE}=(7.32\pm 0.34)\times 10^{-6}\left(\frac{E}{0.451 {\rm GeV}}\right)
^{-2.10\pm0.03} {\rm cm}^{-2}{\rm s}^{-1}{\rm sr}^{-1}{\rm GeV}^{-1}
\end{equation}
between 0.1 and $\sim 50$ GeV (statistics limited) \cite{sre98}.
At energies below 30 MeV, the EGRB spectrum appears to be steeper, as 
determined from an analysis of {\it COMPTEL} data \cite{ka96}.

Figure 1, taken from Ref. \cite{st96}, shows a comparison between the
diffuse inner-galactic and extragalactic spectra measured by {\it EGRET}.
It shows that these diffuse spectra have fundamentally different origins.
The galactic spectrum shows evidence of the predicted ``bump'' from neutral 
pion decay\cite{st77}, \cite{hu97} whereas the extragalactic 
spectrum shows no such feature
as would be expected from cosmic ray $p-p$ interactions. This type of
direct spectral information eliminates purely diffuse extragalactic 
cosmic-ray interaction origin models, such as have been proposed\cite{dar95}
as explanations for the EGRB.

\section{The EGRB from 0.03 to 10 GeV}

The most promising model proposed for the origin of the GeV range 
extragalactic \gray\ background (EGRB), first detected by {\it SAS-2} 
and later confirmed by {\it {\it EGRET}}\cite{sre98},
is that it is the collective emission of an isotropic distribution of faint,
unresolved blazars (See Ref. \cite{ss96} and references therein.).  
Such unresolved blazars are a natural candidate for explaining the EGRB since,
they are the only significant non-burst sources of high energy
extragalactic \grays\ detected by {\it EGRET}.

\begin{figure}
\centerline{\psfig{figure=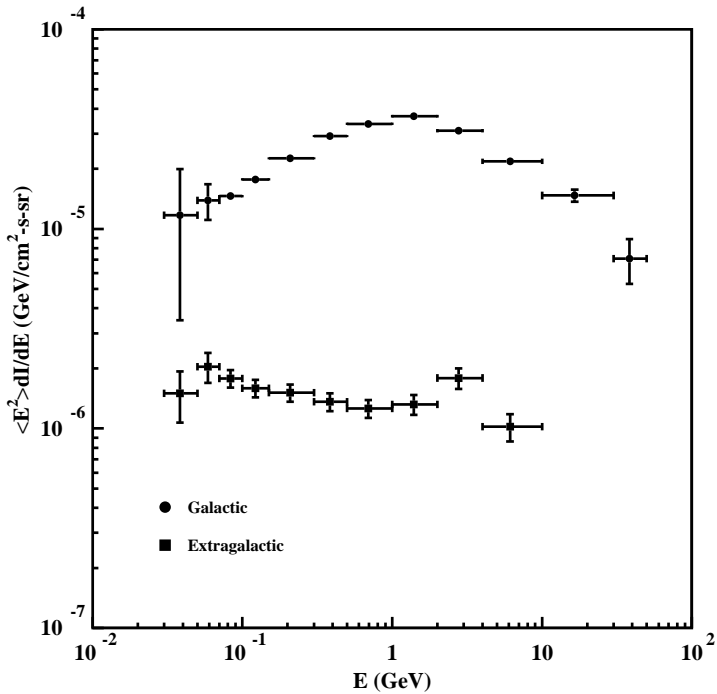,width=4.0in}}
\vspace{-1.0cm}
\caption{Comparison of Galactic and extragalactic diffuse spectra
as determined by {\it EGRET}.}
\end{figure}

\subsection{The Unresolved Blazar Model:}

To determine the collective output of all \gray\ blazars, one can use the 
observed {\it EGRET} distribution of \gray\ luminosities and extrapolate to 
obtain a ``direct'' \gray\ luminosity function (LF) per comoving volume,
$f_{\gamma}(l_{\gamma},z)$ 
\cite{ch95}.
Alternatively, one can make use of much larger catalogs at other wavelengths
and assume a relationship between the source luminosities at the catalog 
wavelength and the GeV region \cite{ssm93}, \cite{pa93}.
Both methods have uncertainties.  

With regard to the former method, only the ``tip of the iceberg'' of the 
\gray\ LF has been observed by {\it EGRET}. Lower luminosity \gray\ sources 
whose fluxes at Earth would  fall below {\it EGRET}'s minimum detectable flux, 
{\it i.e.} {\it EGRET}'s point source sensitivity (PSS), are not detected. 
Extrapolating the \gray\ LF to fainter source luminosities must then involve 
some extra assumption or assumptions.

We have chosen to use the latter method and have assumed a 
linear relation between the luminosities of a source at radio 
and \gray\ wavelengths in an attempt to estimate a LF which would hold
at fainter luminosities. The extent of such a correlation 
is by no means well established \cite{pa93} -- 
\cite{ma97}. 
However, since most theoretical models invoke the same high energy electrons 
as the source of both the radio and \gray\ emission, 
a quasi-linear relation between radio and \gray\ luminosities is a logical 
assumption. In fact, recent observations support this supposition \cite{jo01} .

We used this latter method to estimate the contribution of unresolved
blazars to the EGRB, and found that up to 100\% of the EGRB measured by 
can be accounted for\cite{ss96}. Our
model assumes a linear relationship between the differential \gray\ 
luminosity $l_{\gamma}$ at $E_{f}=0.1$ GeV and the differential radio 
luminosity $l_{r}$ at 2.7 GHz for all sources, $l_{\gamma} \equiv \kappa l_{r}$
with $\kappa$ determined by the observational data.
One can then used the measured radio LF $f_{r}(l_{r},z)$ for blazars 
(flat spectrum radio sources)\cite{dp90} to calculate the collective 
\gray\ output of all blazars. This LF is shown in Figure 2.

\begin{figure}
\centerline{\psfig{figure=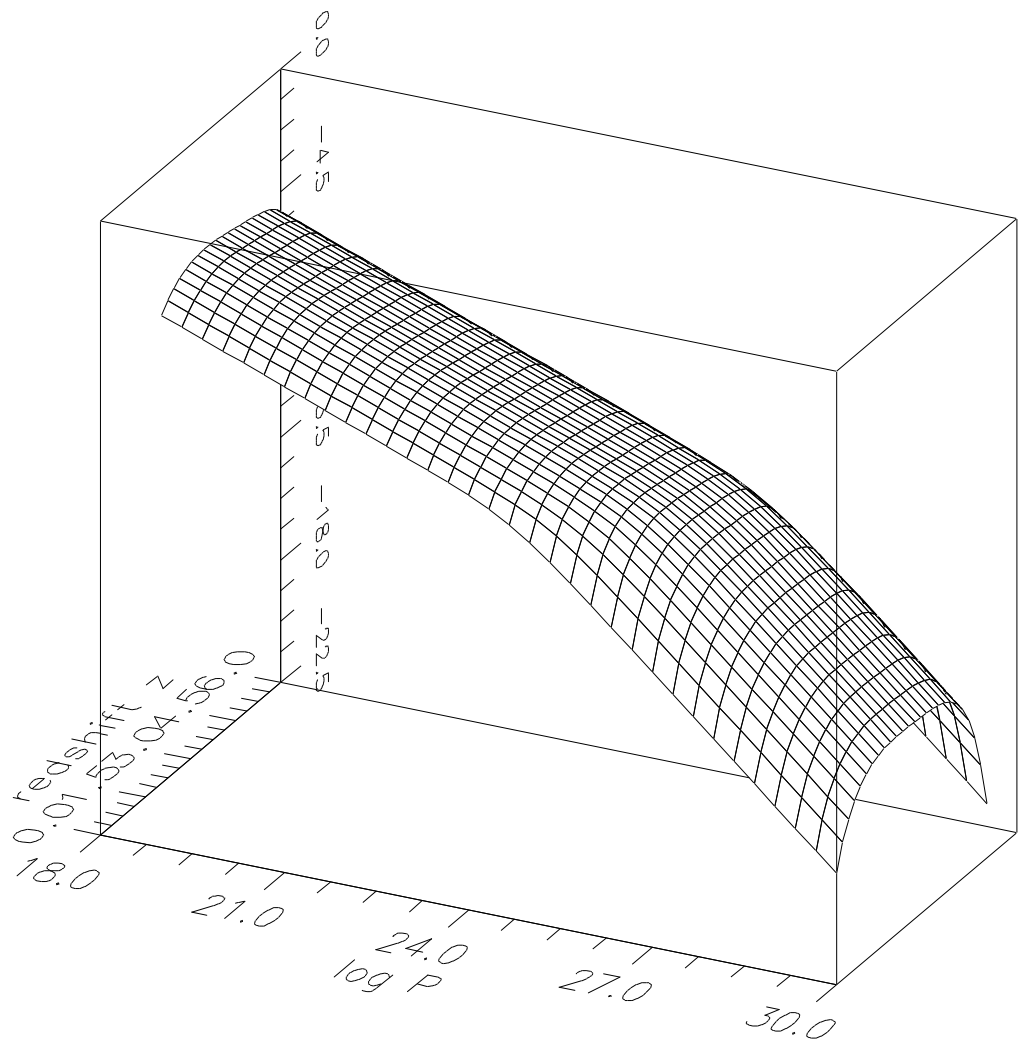,width=4.0in}}
\vspace{8cm}
\caption{Radio luminosity (power) function at 2.7 GHz after Dunlop and
Peacock \protect\cite{dp90}.}
\end{figure}

The simplified elements of our calculations are as follows:
We assume that blazars spend 97\% of their time in a quiescent state
and the remaining 3\% of their time in a flaring state
We assume that the \gray\ and radio LFs in their quiescent
state are related by 
$f_{\gamma}(l_{\gamma},z)=\kappa^{-1}f_{r}(\kappa^{-1}l_{\gamma},z)$.
This relation changes by an average \gray\ ``amplification factor'', 
$\langle A \rangle = 5$, when the blazars are flaring.
We assume that \gray\ spectra for all sources are of the power-law form 
$l(E)=l_{\gamma}(E/E_{f})^{-\alpha}$, where $\alpha$ is assumed to be
independent of redshift. We have taken the distribution of such spectral 
indeces, $\alpha$, from appropriately related {\it EGRET} data. 
We also assume a slight hardening of the
blazar spectra when they are in the flaring state which is supported by
the {\it EGRET} data. For further details, see Ref. \cite{ss96}.

The number of sources ${\cal N}$ detected is a function of the 
detector's PSS at the fiducial energy $E_{f}$, $[F(E_{f})]_{\rm min}$, 
where the integral \gray\ photon flux $F$ is related to $l_{\gamma}$ by

\begin{equation} \label{1.eq}
F(E)=l_{\gamma}(E/E_{f})^{-\alpha}/4\pi\alpha (1+\alpha)^{\alpha +1}R_{0}^{2}r^{2},
\end{equation}

\noindent where $R_{0}r(1+z)$ is the luminosity distance to the source.  
The number of sources at
redshift $z$ seen at Earth with an integral flux $F(E_{f})$ is given by

\begin{equation}
\frac{d{\cal N}}{dF(E_{f})}\Delta F(E_{f})=
\int 4\pi R_{0}^{3}r^{2}\, dr\, f_{\gamma}(l_{\gamma},z(r))\Delta l_{\gamma},
\end{equation}

\noindent where $l_{\gamma}$ in the integrand depends on $z(r)$ and 
$F(E_{f})$ from eq.\ref{1.eq}. The LF, $f_{\gamma}$, includes
both quiescent and flaring terms.
Figure 3 shows the results of our calculation 
of the number of sources versus flux above 0.1 GeV, {\it i.e.}, our predicted 
source count curve,  compared to the {\it EGRET} detections\cite{ss96}. The 
cutoff at $\sim 10^{-7}{\rm cm}^{-2}{\rm s}^{-1}$ for $E_{f}=0.1$ GeV, their 
quoted PSS,
is evident by the dropoff in the detected source count below this flux level.

To calculate the EGRB, we integrate over all sources {\it not} detectable by 
the telescope to obtain the differential number flux of EGRB photons at an 
{\it observed} energy $E_{0}$:

\begin{equation}
\frac{dN_{\gamma}}{dE}(E_{0})=\int 4\pi R_{0}^{3}r^{2}\, dr
\int d\alpha \, p(\alpha)
\int_{l_{min}}^{l_{max}}\frac{dF}{dE}(E_{0}(1+z))f_{\gamma}(l_{\gamma},z)
e^{-\tau(E_{0},z)}\, dl_{\gamma}.
\end{equation}

~
\begin{figure}
\centerline{\psfig{figure=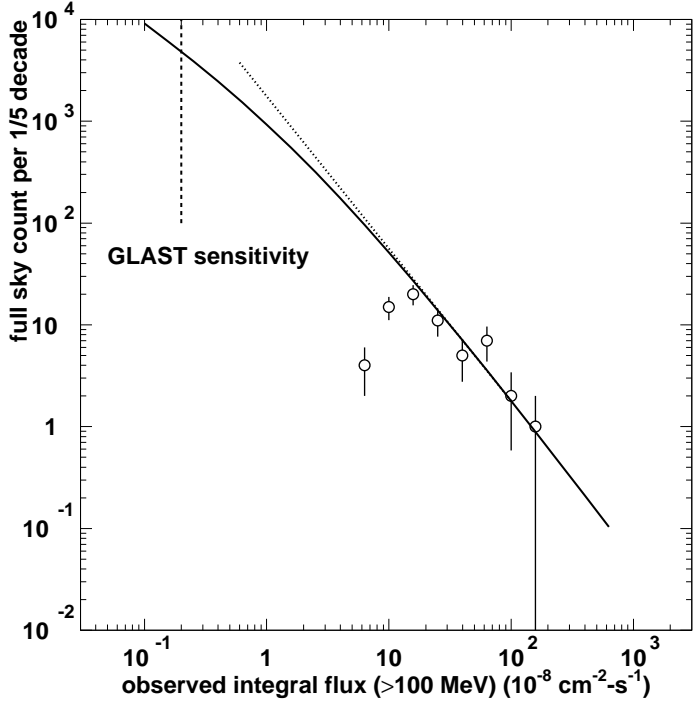,width=4.0in}}
\vspace{-0.1cm}
\caption{Source number count per one-fifth decade of integral flux at Earth
The straight dotted line is the Euclidean relation 
${\cal N}(>F)\propto F^{-3/2}$ for homogeneous distribution of sources.  
The open circles represent the {\it EGRET} blazar
detections and the solid line is the model prediction.}
\end{figure}

\noindent This expression includes an integration over the probability 
distribution of spectral indices $\alpha$ based on the second 
{\it EGRET} Catalog \cite{th95}.

There is also an important attenuation factor in this expression; the
attenuation occurring as the \grays\ produced by blazars
propagate through intergalactic space and interact with 
cosmic UV, optical, and IR background photons to produce $e^{\pm}$ pairs.
If a substantial fraction of the EGRB is from high-$z$ sources, a steepening 
in the spectrum should be seen at energies above $\sim$ 20 GeV caused by the
attenuation effect\cite{ss98}.
Figure 4, from Ref.\cite{ss98}, shows the calculated EGRB spectrum 
(based on the {\it EGRET} PSS) 
compared to {\it EGRET} data.  The slight curvature in the spectrum below 10 
GeV is caused by the distribution of unresolved blazar spectral indeces; the 
harder
sources dominate the higher energy EGRB and the softer sources dominate the
lower energy EGRB. The steepened spectra above $\sim$ 20 GeV in Figure 4 show 
the attenuation effect and its uncertainty.

\begin{figure}
\centerline{\psfig{figure=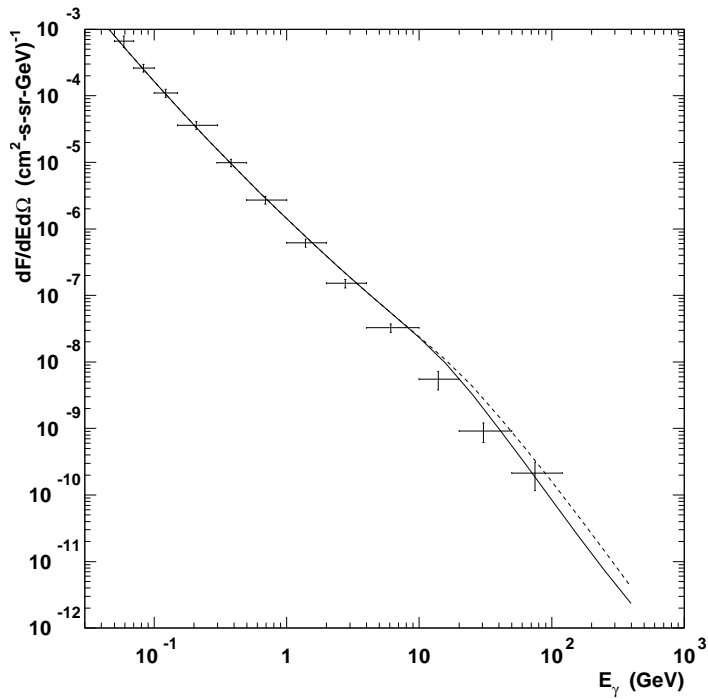,width=4in}}
\caption{The predicted {\it EGRET} EGRB from unresolved blazars compared 
with the {\it EGRET} data. {\it GLAST} should see an EGRB about a factor of 
2 lower at energies above 1 GeV (see text).}
\end{figure}

\subsection{Critique of the Assumption of Independence of Blazar Gamma-Ray
and Radio Luminosities}

Chiang and Mukherjee\cite{cm98} have attempted to calculate the EGRB from
unresolved blazars assuming complete independence between blazar \gray\
and radio luminosities. They then used the intersection between the sets
of flat spectrum radio sources (FSRSs) of fluxes above 1 Jy found in the 
K\"uhr catalogue and the blazars observed by {\it EGRET} as their sample, 
optimizing to the redshift distribution of that intersection set to obtain 
a LF and source redshift evolution. Using this procedure,
they derived a LF which had a low-end cutoff at $10^{46}$ erg s$^{-1}$. 
Then, with no fainter sources included in their analysis, they
concluded that only $\sim$ 1/4 of the 0.1 to 10 GeV EGRB could be accounted 
for as unresolved blazars and that another origin must be found for the EGRB
in this energy range.

We have argued above that it is reasonable to expect that the radio and \gray\
luminosities of blazars are correlated. Any such correlation will destroy the
assumption of statisitical independence made by Chiang and Mukherjee and 
introduce a bias in their analysis. In fact, their analysis leads to many
inconsistencies. Among them are the following:

A. The LF derived by Chiang and Mukherjee\cite{cm98} allows for no sources 
with luminosities below
$10^{46}$ erg s$^{-1}$. In fact, {\it all} of the six sources found by
{\it EGRET} at redshifts below $\sim$ 0.2 have luminosites between
$\sim10^{45}$ erg s$^{-1}$ and $\sim10^{46}$ erg s$^{-1}$\cite{mu01}. 
Elimination of fainter sources from the analysis can only lead to a lower 
limit on the EGRB from unresolved blazars. The fainter sources contribute
significantly in acounting for unresolved blazars being the dominant 
component of the EGRB. (In this regard, see also, Ref.\cite{mp00}.)

B. Chiang and Mukherjee limit the {\it EGRET} sources in their analysis
only to the FSRSs in the K\"uhr catalogue. However, if there is truly no 
correlation between blazar radio and \gray\ luminosities, then any of the 
millions
of FSRSs given by the Dunlop and Peacock radio LF \cite{dp90} are equally
likely to be {\it EGRET} sources. In that case, of the 50 odd sources in the
2nd {\it EGRET} catalogue, virtually {\it none}, {\it i.e.} $\sim 10^{-6}$,
should be K\"uhr sources.

The above discussion indicates that the assumption of non-correlation between
the radio and \gray\ fluxes of blazars made by Chiang and Mukherjee in their
analysis is not a good one and that this assumption invalidates their 
conclusions.

\subsection{{\it GLAST} and the EGRB:}

With an estimated point 
source sensitivity (PSS) nearly two orders of magnitude lower than {\it EGRET}'s, {\it GLAST} will be able
to detect $\cal O$$(10^{2})$ times more blazars than {\it EGRET}, and measure the EGRB spectrum to $>1$ TeV (assuming the {\it EGRET} power law spectrum).  
These two capabilities will enable {\it GLAST} to either strongly support or 
reject the unresolved-blazar hypothesis for the origin of the EGRB.

Figure 3 shows that $\cal O$$(10^{3})$ blazars should be detectable by 
{\it GLAST}, assuming it achieves a PSS of $\sim 2\times 10^{-9}$ 
cm$^{-2}$s$^{-1}$. Using this PSS and our derived source count curve
as shown in Figure 3, we have estimated that the remaining
``diffuse'' EGRB seen by {\it GLAST} should be a factor of $\sim 2$ lower
for $E>1$ GeV. Below 1 GeV, this factor of 2 will not apply because source 
confusion owing to the poorer angular resolution of {\it GLAST} at these lower
energies will reduce the number of blazars resolved out of the background.

We conclude that {\it GLAST} can test the unresolved blazar background model 
in three ways: 

A. {\it GLAST} should see roughly 2 orders of magnitude more blazars than 
{\it EGRET} because of its ability to detect the fainter blazars which 
contribute to the EGRB in our model. It can thus make a much deeper 
determination of the source count curve. {\it GLAST} can also
determine the redshift distribution of many more identified \gray\
blazars, using its better point source angular resolution to make 
identifications with optical sources having measured redshifts.  
With its larger dynamic range, {\it GLAST} can then test the assumption 
of an average linear relation between 
the \gray\ and radio fluxes of identified blazars. 
All of these determinations will test the basic assumptions and results of our 
model.

B. With its better PSS, {\it GLAST} will resolve out more blazars 
from the background. Thus, fewer unresolved blazars will be left to
contribute to the EGRB.  reducing the level of the measured EGRB 
compared to {\it EGRET}'s by a factor of $\sim$2 if our predictions are
correct.

C. The much greater aperture of {\it GLAST} at 100 GeV will allow a
determination of whether or not a steepening exists in the EGRB, since
the number of EGRB \grays\ recorded by {\it GLAST} above 100 GeV 
will be of order $10^{3}$ to $10^{4}$, assuming a
continuation of the {\it EGRET} power-law spectrum. Such a steepening
can be caused by both absorption and intrinsic turnovers in blazar spectra.
Given enough sub-TeV spectra of individual blazars with known redshifts, 
these two effects can be separated.

\section{The EGRB Between 0.5 and 30 MeV}

The explanation for the origin of the EGRB at energies in the range of several
MeV must be a non-blazar explanation. The reason for this is that while the 
EGRB spectrum in this energy range appears to be {\it softer} than that at 
higher energies \cite{ka96}, the data from {\it OSSE} and {\it COMPTEL} on 
individual blazars in this energy range indicate a {\it harder} spectrum than 
that at higher energies. The measured blazar spectra appear to break below 
$\sim$ 10 MeV to spectra with a typical power-law index of $\sim$ 1.7 
\cite{mc95}. Thus, even if unresolved blazars account for almost all of the 
EGRB in the 0.1 to 10 GeV range, this cannot be the case at lower energies.

Calculations have shown that a superposition of redshifted lines from Type Ia 
and Type II supernovae should reasonably provide a significant component of
the EGRB at energies $\sim$ 1 MeV. The important line emission is from the 
decay chain $^{56}$Ni $\rightarrow$ $^{56}$Co $\rightarrow$ $^56$Fe and also
from the decay of $^{26}$Al, $^{44}$Ti and $^{60}$Co \cite{th93}, \cite{wa99}.
However, supernovae cannot account for the entire EGRB in this energy range,
since they produce no line emission above 3.5 MeV.

Another serious possibility as a significant contributer to the multi-MeV
EGRB is non-thermal tails in the energy spectra of the AGN\cite{st99}.
These would be the same AGN which have recently been resolved out by the 
{\it Chandra} telescope and found to be the dominant component of the once
unresolved X-ray background\cite{mu00}. 

A recent discussion of AGN models fitting the X-ray background has been
given in Ref. \cite{gi01}. While there are no data on individual AGN in the
multi-MeV energy range at the present time, Stecker, Salamon and Done 
\cite{st99} have pointed to the galactic black hole candidate 
Cyg. X-1 as an example of a black hole source which has been 
shown from {\it COMPTEL} data to have a non-thermal tail extending to 
multi-MeV energies\cite{mc00}. If the extragalactic black
hole sources which make up the X-ray background have such non-thermal tails,
they may account for most of the EGRB in the multi-MeV range.

It should be noted that the extraction of the $\sim$ MeV EGRB from the raw 
{\it COMPTEL} data is a difficult process, in part owing to the fact that this
double Compton scattering telescope was not designed to measure this
background. In our opinion, a dedicated low-mass, free flyer satellite, 
specifically designed to measure the EGRB at low \gray\ energies will be 
required in order to accurately determine its characteristics.

\section{The EGRB Above 10 GeV}

It has already been pointed out that the EGRB should break above $\sim$ 20 GeV
energy owing to absorption of high energy \grays\ by pair-production 
interactions with lower energy starlight photons \cite{ss98}.
There is also another potential cause for a steepening in the EGRB 
from blazars. The {\it EGRET} detector obtained rough power-law spectral 
indeces for blazars in the 0.1 to 10 GeV energy decade, however, we presently
have no data for these objects in the 10 to 100 GeV decade.
Presently popular theoretical models
predict that the spectra of highly luminous blazars will exhibit a cutoff at
energies in the 10 to 100 GeV range, whereas the less luminous X-ray selected
BL Lac objects can have spectra extending into the TeV energy range
\cite{sds96},\cite{fo98}. 

Indeed, there have now been ground based detections of at least 5 X-ray 
selected BL Lac objects (Weekes, these proceedings),  
some of whose spectra extend to multi-TeV energies. While no other types of
blazars have been seen at TeV energies, this may be an result of intergalactic
\gray\ absorption\cite{ss98}, \cite{st92}, \cite{sd98} so that we do not 
really know if 
their intrinsic spectra turn down at energies in the 10 to 100 GeV decade. 
The {\it GLAST} telescope should provide this knowledge in the 
not-too-distant future.  

If the spectra of most blazars possess intrinsic cutoffs above 10 GeV, then
the EGRB from unresolved blazars would be expected to turn over as well.
This effect should be more dramatic than the steepening in the EGRB predicted
from the effect of intergalactic absorption\cite{ss98}. In that case,
if the {\it EGRET} results on the EGRB up to 100 GeV are correct, a new 
component may be present in this higher energy range. Such a component
has been predicted to be produced by the decay of $\sim$ TeV mass higgs 
bosons from cosmic string processes in flat-potential supersymmetric
models\cite{bss98}. Of course, there may be other unknown possibilities as
well.

\section{Conclusions}

We have a workable and testable hypothesis for the origin of the extragalactic
\gray\ background measured by {\it EGRET}, {\it viz.}, that it is made up
primarily of unresolved blazars. The {\it GLAST} \gray\ telescope, to be
flown in the near future, will be able to test this hypothesis in three
ways, {\it i.e.}, (a) by potentially resolving out and detecting thousands of 
more sources, (b) by measuring the remaining background flux, and (c) by
determining the shape of the EGRB up to TeV energies. The many new 
ground-based detectors now under construction will supplement
this information by discovering new extragalactic sources of \grays\ of 
energies above 50 GeV.

On the other hand, the mystery of the origin of the EGRB in the MeV energy
range must await a better determination of this background by a future
dedicated satellite detector.

\end{document}